\documentclass[aps,prl,twocolumn,groupedaddress,showpacs]{revtex4}

\usepackage{graphicx}
\usepackage{dcolumn}
\usepackage{bm}

\begin{document}

% Use the \preprint command to place your local institutional report
% number in the upper righthand corner of the title page in preprint mode.
% Multiple \preprint commands are allowed.
% Use the 'preprintnumbers' class option to override journal defaults
% to display numbers if necessary
%\preprint{}

%Title of paper
\title{Can hadronic rescattering explain the large elliptic flow and 
small HBT radii \\seen at RHIC?}

% repeat the \author .. \affiliation  etc. as needed
% \email, \thanks, \homepage, \altaffiliation all apply to the current
% author. Explanatory text should go in the []'s, actual e-mail
% address or url should go in the {}'s for \email and \homepage.
% Please use the appropriate macro foreach each type of information

% \affiliation command applies to all authors since the last
% \affiliation command. The \affiliation command should follow the
% other information
% \affiliation can be followed by \email, \homepage, \thanks as well.
\author{T. J. Humanic}
\email[]{humanic@mps.ohio-state.edu}
\homepage[]{http://vdgus1.mps.ohio-state.edu/}
%\thanks{}
%\altaffiliation{}
\affiliation{Department of Physics, The Ohio State University,
Columbus, OH 43210}

%Collaboration name if desired (requires use of superscriptaddress
%option in \documentclass). \noaffiliation is required (may also be
%used with the \author command).
%\collaboration can be followed by \email, \homepage, \thanks as well.
%\collaboration{}
%\noaffiliation

\date{\today}

\begin{abstract}
Results from the data obtained in the first physics run of the
Relativistic Heavy Ion Collider (RHIC)
have shown suprisingly large elliptic flow and suprisingly small
HBT radii. Attempts to explain both results in a consistant
picture have so far been unsuccessful. The present work shows 
that a thermal model + hadronic rescattering 
calculation can explain both elliptic flow and HBT 
results from RHIC. The calculation 
requires a very early
hadronization time of about 1 fm/c after the initial
collision of the nuclei.
\end{abstract}

% insert suggested PACS numbers in braces on next line
\pacs{25.75.Gz, 25.75.Ld}
% insert suggested keywords - APS authors don't need to do this
%\keywords{}

%\maketitle must follow title, authors, abstract, \pacs, and \keywords
\maketitle

% body of paper here - Use proper section commands
% References should be done using the \cite, \ref, and \label commands
%\section{}
% Put \label in argument of \section for cross-referencing
%\section{\label{}}
%\subsection{}
%\subsubsection{}

Results of the Year-1 running of the Relativistic Heavy Ion Collider
(RHIC) for Au+Au collisions at $\sqrt{s}=130$ GeV have shown 
suprisingly large pion elliptic flow \cite{Adler:2001a} and 
suprisingly small
radii from two-pion Hanbury-Brown-Twiss interferometry (HBT)
\cite{Adler:2001b,Adcox:2002a}. 
Attempts to explain both results 
in a consistant picture have so far been unsuccessful. Hydrodynamical
models agree with the large elliptic flow seen in the RHIC data 
\cite{Kolb:1999a} but significantly disagree with the experimental HBT
radii \cite{Rischke:1996a}. On the other hand, relativistic quantum
molecular dynamics calculations which include hadronic 
rescattering, for example RQMD v2.4 \cite{Sorge:1989a},
significantly underpredict the elliptic flow seen in the RHIC data
\cite{Sorge:1995a} but predict pion HBT radii comparable to the data 
\cite{Hardtke:1999a}.
A calculation has recently been made to extract HBT radii 
with a hydrodynamical
model coupled with a hadronic rescattering afterburner with the result
that the HBT radii are significanly larger than measurements
\cite{Soff:2000a}. This lack of a single model to explain both results
has been our first big mystery from RHIC. It has been
suggested that we should call into question 
our current understanding of what information 
pion HBT measurements give us \cite{Gyulassy:2001a}.

In an effort to address this mystery, the present work explores
a somewhat different picture of the nuclear collision than those
presented above. In this picture, hadronization into a thermal equilibrium
state occurs soon (about 1 fm/c) after the initial collision of the
nuclei followed by hadronic rescattering until freezeout. The goal
will thus be to test whether hadronic rescattering alone
can generate enough general flow in the system to explain both 
the elliptic flow and HBT results from RHIC. Note that this approach,
while similar to the relativistic quantum molecular dynamics
calculations mentioned above, differs from them in the choice of 
the model of the initial state of the system before hadronic 
rescattering commences (e.g. the other models use a color string
picture for the initial state) \cite{Sorge:1989a}. The price for
using the present method is that hadronic-like objects must
exist during the high-energy-density ($\rho>1$ $GeV/fm^3$) 
phase of the collision. The discussion of
this important point is deferred until later.

A brief description of the rescattering model
calculational method is given
below. The method used is similar to that used in previous 
calculations for lower CERN Super Proton Synchrotron (SPS)
energies \cite{Humanic:1998a}. 
Rescattering is simulated with a semi-classical 
Monte Carlo calculation which assumes strong binary collisions 
between hadrons. The Monte Carlo calculation is
carried out in three stages: 1) initialization and hadronization, 2)
rescattering and freeze out, and 3) calculation of experimental 
observables. Relativistic kinematics is used 
throughout.  All calculations are made to simulate RHIC-energy
Au+Au collisions in order to compare with the results of the
Year-1 RHIC data.

The hadronization model employs simple parameterizations to describe the 
initial momenta and space-time of the hadrons similar to
that used by Herrmann and Bertsch \cite{Herrmann:1995a}. The initial 
momenta are assumed to follow a thermal transverse
(perpendicular to the beam direction)
momentum distribution for all particles,
\begin{equation}
(1/{m_T})dN/d{m_T}=C{m_T}/[\exp{({m_T}/T)} \pm 1]
\end{equation}
where ${m_T}=\sqrt{{p_T}^2 + {m_0}^2}$ is the transverse mass, $p_T$ 
is the transverse momentum, $m_0$ is the particle rest mass, $C$ is 
a normalization constant, and $T$ is the initial temperature
of the system, 
and a gaussian rapidity distribution for mesons,
\begin{equation}
dN/dy=D \exp{[-{(y-y_0)}^2/(2{\sigma_y}^2)]}
\end{equation}
where $y=0.5\ln{[(E+p_z)/(E-p_z)]}$ is the rapidity, $E$ is the 
particle energy, 
$p_z$ is the longitudinal (along the beam direction)
momentum, $D$ is a normalization constant, 
$y_0$ is the central
rapidity value (mid-rapidity), and $\sigma_y$ is the rapidity width.
Two rapidity distributions for baryons have been tried: 1) flat
and then falling off near beam rapidity and 2) peaked at central
rapidity and falling off until beam rapidity. Both baryon
distributions give about the same results. 
The initial space-time of the
hadrons for $b=0$ fm (i.e. zero impact parameter or central collisions) 
is parameterized as having cylindrical symmetry with 
respect to the 
beam axis. The transverse particle density dependence is assumed 
to be that of a
projected uniform sphere of radius equal to the projectile radius, $R$ 
($R={r_0}A^{1/3}$, where ${r_0}=1.12$ fm and $A$ is the
atomic mass number 
of the projectile). For $b>0$ (non-central collisions) the transverse
particle density is that of overlapping projected spheres.
The longitudinal
particle hadronization position ($z_{had}$) and time ($t_{had}$) 
are determined by the relativistic equations \cite{Bjorken:1983a},
\begin{equation}
z_{had}=\tau_{had}\sinh{y};   t_{had}=\tau_{had}\cosh{y}
\end{equation}
where $y$ is the particle rapidity and $\tau_{had}$ is the 
hadronization proper time.
Thus, apart from particle multiplicities, the hadronization model has 
three free
parameters to extract from experiment: $\sigma_y$,
$T$ and $\tau_{had}$.
The hadrons included in the calculation are pions, kaons,
nucleons and lambdas
($\pi$, K, N, and $\Lambda$), and the $\rho$, $\omega$, $\eta$, 
${\eta}'$, 
$\phi$, $\Delta$, and $K^*$ resonances. For simplicity, the
calculation is isospin averaged (e.g. no distinction is made among
a $\pi^{+}$, $\pi^0$, and $\pi^{-}$). Resonances are present at 
hadronization
and also can be produced as a result of rescattering. Initial resonance
multiplicity fractions are taken from 
Herrmann and Bertsch \cite{Herrmann:1995a}, 
who extracted results from the HELIOS experiment \cite{Goerlach:1992a}. 
The initial resonance fractions used in
the present calculations are: $\eta/\pi=0.05$, $\rho/\pi=0.1$, 
$\rho/\omega=3$, $\phi/(\rho+\omega)=0.12$,
${\eta}'/\eta=K^*/\omega=1$ and, for simplicity, $\Delta/N=0$.

The second stage in the calculation is rescattering 
which finishes with the
freeze out and decay of all particles. Starting 
from the initial stage ($t=0$ fm/c), the positions 
of all particles are allowed to evolve in time in small
time steps ($dt=0.1$ fm/c) according to their 
initial momenta. At each time step
each particle is checked to see a) if it decays, and b) if it is 
sufficiently
close to another particle to scatter with it.
Isospin-averaged s-wave and p-wave cross sections 
for meson scattering are 
obtained from Prakash et al.
\cite{Prakash:1993a}. The calculation is carried out to 100 fm/c,
although most of the rescattering finishes by about 30 fm/c.
The rescattering calculation
is described in more detail elsewhere \cite{Humanic:1998a}.

Calculations are carried out assuming initial parameter values and particle
multiplicities for each type of particle. In the last stage of the 
calculation, the freeze-out and decay momenta and 
space-times are used to produce
observables such as pion, kaon, and nucleon
multiplicities and transverse momentum and rapidity distributions. 
The values of the 
initial parameters of the calculation and multiplicities
are constrained to give observables which agree with 
available measured hadronic observables. As a cross-check
on this, the total kinetic energy from the calculation is 
determined and
compared with the RHIC center of mass energy of 
$\sqrt{s}=130$ GeV to see that they are in 
reasonable agreement. Particle multiplicities were estimated from
the charged hadron multiplicity measurements of the RHIC 
PHOBOS experiment \cite{Back:2000a}. Calculations were carried
out using isospin-summed events containing at
freezeout about 5000 pions, 500 kaons, and 650 nucleons.
The hadronization model parameters used were $T=300$ MeV,
$\sigma_y$=2.4, and $\tau_{had}$=1 fm/c. It is interesing
to note that the same value of $\tau_{had}$ was required 
in a previous rescattering calculation to successfully 
describe results from SPS
Pb+Pb collisions \cite{Humanic:1998a}. Figure 1 shows
$m_T$ distributions for pions, kaons, and nucleons from the
rescattering calculation for $b=0$ fm near midrapidity ($-1<y<1$)
fitted to exponentials
of the form $exp{(-m_T/B)}$, where $B$ is the slope parameter.
\begin{figure}
\includegraphics{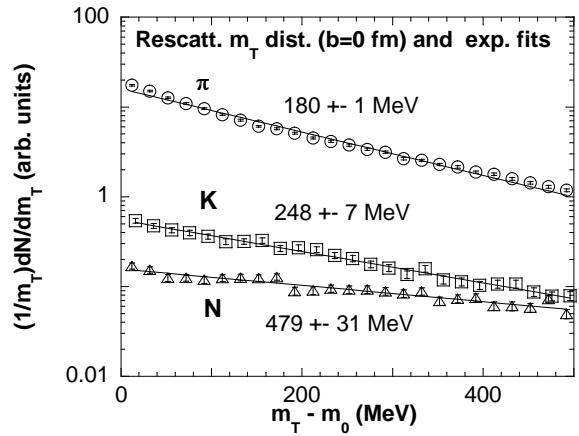}
\caption{\label{fig:mT} Transverse mass distributions from the
rescattering model. The lines are exponential fits to the
distributions and the slope parameters are shown.}
\end{figure}
The extracted slope parameters shown in Figure 1 are close in value
to preliminary measurements from the STAR experiment for the
$\pi^-$, $K^-$, and anti-proton of $190\pm10$, 
$300\pm30$,
and $565\pm50$ MeV, respectively \cite{Adler:2002a}. Thus, we see that
if all hadrons begin at a common temperature of 300 MeV,
the hadronic rescattering alone is able to generate enough radial
flow to account for the differences in slope among the 
pion, kaon, and nucleon $m_T$ distributions.

The elliptic flow 
and two-pion HBT observables
are also calculated from the freeze-out
momenta and space-time positions 
of the particles at the end of the rescattering stage.
The elliptic flow variable, $v_2$, is defined 
as \cite{Poskanzer:1998a}
\begin{equation}
v_2 = \langle cos(2\phi) \rangle;   \phi = \arctan(p_y/p_x)
\end{equation}
where $p_x$ and $p_y$ are the $x$ and $y$ components of the particle
momentum, and $x$ is in the impact parameter direction and
$y$ is in the ``out of plane'' direction (i.e. $x-z$ is the
reaction plane and $z$ is the beam direction).
The HBT pion source parameters are extracted from the
rescattering calculation using the same method as was applied
for previous SPS-energy rescattering calculations \cite{Humanic:1998a}.
The Pratt-Bertsch ``out-side-long'' radius parameterization is used
\cite{Pratt:1990a,Bertsch:1989a} yielding the four parameters
$R_{Tside}$, $R_{Tout}$, $R_{Long}$, and $\lambda$, which
represent two mutually perpendicular transverse 
(to the beam direction) radius parameters,
a radius parameter along the beam direction, and
a parameter related to the ``strength'' of the two-pion 
correlations, respectively.

Figure 2 shows the $p_T$ dependence of $v_2$ for pions and
nucleons extracted from the $b=8$ fm rescattering calculation
compared with the trends of the STAR 
measurements for $\pi^{+}+\pi^{-}$ and $p$ + $p-bar$ 
at $11-45\%$ centrality \cite{Adler:2001a}, which 
roughly corresponds to this impact parameter. 
\begin{figure}
\includegraphics{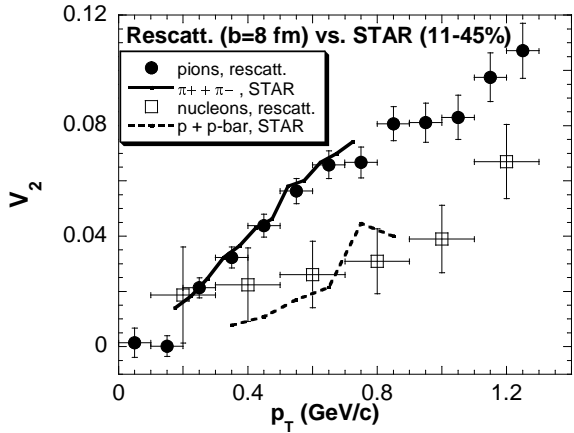}
\caption{\label{fig:v2} Comparison of $v_2$ calculated from the
rescattering model for $b=8$ fm with STAR measurements for pions
and nucleons. The plotted points with error bars are the
rescattering calculations and the lines show the trends of the
STAR measurements. Average errors on the STAR measurements
are $\leq0.002$ for pions and $0.006$ for protons+antiprotons.}
\end{figure}
As seen, the 
rescattering calculation values
are in reasonable agreement with the STAR measurements. 
Thus, the same
rescattering mechanism that can account for the radial flow seen
in Figure 1 also can account for the magnitude and $p_T$ dependence
of the elliptic flow for pions and nucleons.

The pion source parameters extracted from HBT analyses of
rescattering calculations for three different impact
parameters, $b=0$, $5$, and $8$ fm, are compared with 
STAR $\pi^{-}$ measurements at three
centrality bins \cite{Adler:2001b} in Figure 3.
\begin{figure*}
\includegraphics{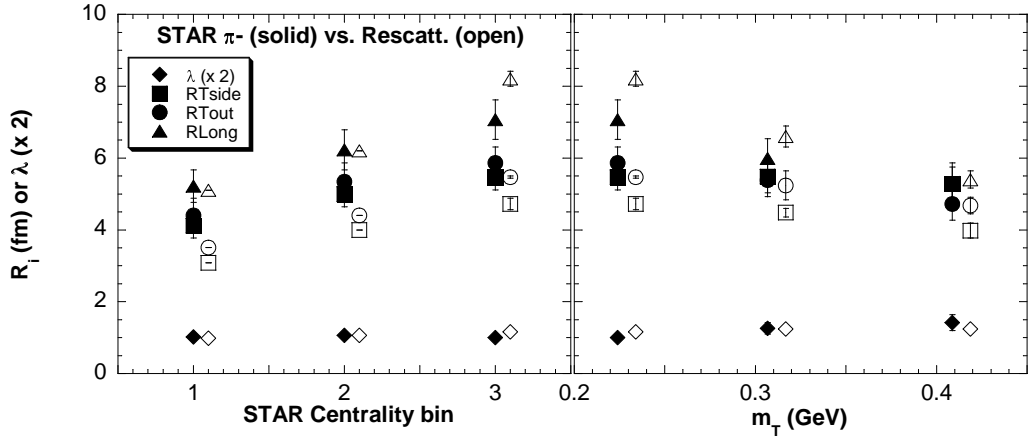}
\caption{\label{fig:hbt} Comparison of HBT source parameters from
rescattering with STAR measurements as a function of centrality
bin (see text) and $m_{T}$. The STAR measurements are the solid
symbols and the rescattering calculations are the open symbols.
The errors on the STAR measurements are statistical+systematic.}
\end{figure*}
Note that the PHENIX HBT results \cite{Adcox:2002a} 
are in basic agreement with the STAR results. 
The STAR
centrality bins labeled ``3'', ``2'', and ``1'' in the figure
correspond to $12\%$ of central, the next $20\%$, and the 
next $40\%$,
respectively. These bins are roughly approximated by the impact
parameters used in the rescattering calculations,
i.e. the average impact parameters of the STAR centrality
bins are estimated to be within $\pm2$ fm of the
rescattering calculation impact parameters used to
compare with them.
In the left
panel, the centrality dependence of the HBT parameters
is plotted for a $p_T$ bin of $0.125-0.225$ GeV/c. In the right
panel, the $m_T$ dependence of the HBT parameters is plotted
for centrality bin 3, for the STAR measurments, or $b=0$ fm,
for the rescattering calculations. Although there are
differences in some of the details, the trends of the STAR
HBT measurements are seen to be described rather well by the
rescattering calculation.

As shown above, the radial and elliptic flow as well 
as the features of the HBT 
measurements at RHIC can be adequately described by
the rescattering model with the hadronization model parameters
given earlier. The results of the calculations are found
to be sensitive to the value of $\tau_{had}$ used, as was
studied in detail for SPS rescattering calculations 
\cite{Humanic:1998a}. For calculations with $\tau_{had}>1$ fm/c
the initial hadron density is smaller, fewer collisions occur, 
and the rescattering-generated flow is reduced, reducing
in magnitude the radial and elliptic flow and most 
of the HBT observables. Only the HBT parameter $R_{Long}$
increases for larger $\tau_{had}$ reflecting the
increased longitudinal size of the initial hadron source,
as seen in Equation 3. One can compensate for this
reduced flow in the other observables by introducing an
ad hoc initial ``flow velocity parameter'', but the increased 
$R_{Long}$ cannot be compensated by this new parameter.
In this sense, the initial hadron model used in the
present calculations with $\tau_{had}\sim1$ fm/c
and no initial flow is uniquely determined with
the help of $R_{Long}$.

At this point, it is appropriate to discuss
how physical the initial conditions
of the present rescattering calculation are.
In order to use this picture, one must assume: 1) hadronization 
occurs very rapidly
after the nuclei have passed through each other, i.e.
$\tau_{had}=1$ fm/c, 2) the hadrons thermalize rapidly,
and 3) hadrons or at least hadron-like objects can exist
in the early stage of the collision where there are maximum values of
$T$ and $\rho$ of 300 MeV 
and 8 GeV/$fm^3$, respectively.

Addressing assumption 3) first, in the calculation the maximum 
number density
of hadrons at mid-rapidity at $t=0$ fm/c is 6.8 $fm^{-3}$, 
rapidly dropping to about
1 $fm^{-3}$ at $t=4$ fm/c. Since most of these hadrons are pions,
it is useful as a comparison to estimate the effective volume of a
pion in the context of the $\pi-\pi$ scattering cross section,
which is about 0.8 $fm^2$ for s-waves \cite{Prakash:1993a}.
The ``radius'' of a pion is found to be 0.25 fm and
the effective pion volume is 0.065 $fm^3$, the reciprocal
of which is about 15 $fm^{-3}$. From this it is seen that at the 
maximum hadron number
density in the calculation, the particle occupancy 
of space is estimated to be
less than $50\%$, falling rapidly with time. One could
speculate that this may
be enough spacial separation to allow individual hadrons or
hadron-like objects to keep 
their identities and not melt into quark matter,
resulting in a ``super-heated'' semi-classical gas
of hadrons at very early times, as assumed in the
present calculation.

Assumptions 1) and 2) can both be motivated by the Color Glass
Condensate model \cite{McLerran:2002a,Kovner:1995a}.
Before the collision of two relativistic nuclei, the
nuclei can be described as two thin (lorentz-contracted)
sheets of Color Glass (a dense glass of gluons). The use of
the term ``glass'' is not merely an analogy but is mathematically
rigorous due to the time-dilated nature of the gluon field
which behaves like a liquid on long time scales but a
solid on short ones \cite{McLerran:2002a}. In the usual
picture, after the collision
takes place the Color Glass melts into quarks and gluons in
a timescale of about 0.3 fm/c at RHIC energy, and then the
matter expands and thermalizes into quark matter by about 
1 fm/c. The timescale
in the Color Glass Condensate model for thermalized
matter matches the timescale needed in the rescattering
model for an initial thermalized hadron gas. Thus
one is tempted to modify the collision senario such that 
instead of the
Color Glass melting into quarks and gluons just after
the collision, the sudden impact of the collision
``shatters'' it
directly into hadronic
fragments which then thermalize on the same timescale
as in the parton senario due to the hadronic 
strong interactions.

In summary, a thermal model + hadronic rescattering picture
is able to adequately describe the large elliptic flow and 
small HBT radii recently measured at RHIC. A feature of this
picture is a very early hadronization time of about 1 fm/c
after the initial collision of the nuclei.

% If you have acknowledgments, this puts in the proper section head.
\begin{acknowledgments}
It is a pleasure to acknowledge Ulrich Heinz and Mike Lisa
for their helpful
suggestions regarding this work and Larry McLerran 
for illuminating discussions 
on the Color Glass 
Condensate model. This work was supported by the U.S.
National Science Foundation under grant PHY-0099476.
\end{acknowledgments}

% Create the reference section using BibTeX:
%\bibliography{basename of .bib file}

% figures should be put into the text as floats.
% Use the graphics or graphicx packages (distributed with LaTeX2e)
% and the \includegraphics macro defined in those packages.
% See the LaTeX Graphics Companion by Michel Goosens, Sebastian Rahtz,
% and Frank Mittelbach for instance.
%
% Here is an example of the general form of a figure:
% Fill in the caption in the braces of the \caption{} command. Put the label
% that you will use with \ref{} command in the braces of the \label{} command.
% Use the figure* environment if the figure should span across the
% entire page. There is no need to do explicit centering.

% Surround figure environment with turnpage environment for landscape
% figure
% \begin{turnpage}
% \begin{figure}
% \includegraphics{}%
% \caption{\label{}}
% \end{figure}
% \end{turnpage}

\end{document}